\documentclass[review,1p,times]{elsarticle}
\usepackage{epsfig}
\usepackage{graphicx}
\usepackage{amsmath}
\usepackage{amssymb}
\usepackage{amsfonts}
\usepackage{tabularx}
\usepackage{color}
\usepackage{dcolumn}% Align table columns on decimal point
\usepackage{bm}% bold math

\begin{document}

\title{{\color{black}Reciprocal and real space maps for EMCD experiments}}

\author[elmin]{Hans Lidbaum}
\author[fysik,cz]{J\'{a}n Rusz}
\author[elmin]{Stefano Rubino}
\author[fysik,andreas]{Andreas Liebig}
\author[fysik]{Bj\"{o}rgvin Hj\"{o}rvarsson}
\author[fysik]{Peter M. Oppeneer}
\author[fysik]{Olle Eriksson}
\author[elmin]{Klaus Leifer}
\ead{klaus.leifer@angstrom.uu.se}

\address[elmin]{Department of Engineering Sciences, Uppsala University, Box 534, S-751 21 Uppsala, Sweden}
\address[fysik]{Department of Physics and Materials Science, Uppsala University, Box 530, S-751 21 Uppsala, Sweden}
\address[cz]{Institute of Physics, Academy of Sciences of the Czech Republic, Na Slovance 2, CZ-182 21 Prague, Czech Republic}
\address[andreas]{Present address: Institute of Physics, Chemnitz University of Technology, D-09107 Chemnitz, Germany}

\date{\today}

\begin{abstract}
Electron magnetic chiral dichroism (EMCD) is an emerging tool for quantitative measurements of magnetic properties using the transmission electron microscope (TEM), with the possibility of nanometer resolution. The geometrical conditions, data treatment and electron gun settings are found to influence the EMCD signal. In this article, particular care is taken to obtain a reliable quantitative measurement of the ratio of orbital to spin magnetic moment using energy filtered diffraction patterns. For this purpose, we describe a method for data treatment, normalization and selection of mirror axis. The experimental results are supported by theoretical simulations based on dynamical diffraction and density functional theory. Special settings of the electron gun, so called telefocus mode, enable a higher intensity of the electron beam, as well as a reduction of the influence from artifacts on the signal. Using these settings, we demonstrate the principle of acquiring real space maps of the EMCD signal. This enables advanced characterization of magnetic materials with superior spatial resolution.
\end{abstract}

%\pacs{68.37.Lp, 75.70.Ak, 78.20.Bh, 79.20.Uv}
\maketitle

%*******************************************************************
%*******************************************************************
%**********************  INTRODUCTION  *****************************
%*******************************************************************
\section{Introduction}
The discovery of the electron magnetic chiral dichroism (EMCD) in the electron energy-loss spectra (EELS) \cite{nature, thesisStefano} demonstrated the feasibility of atom specific magnetic measurements with the transmission electron microscope (TEM). EMCD probes the same electronic transitions as its older x-ray counterpart, x-ray magnetic circular dichroism (XMCD) \cite{stohr} but with a theoretical spatial resolution one or two orders of magnitude better than XMCD. The dependence of the EMCD signal on the specific scattering geometry means that, at best, only qualitative information could be obtained by experiments alone, such as relative changes of local atom specific magnetization \cite{RubinoJMR}. The development of a density functional theory (DFT) based simulation software \cite{prbtheory} and the recent derivations of the sum rule expressions for EMCD \cite{oursr,lionelsr} provide a basis for using the EMCD technique for quantitative measurements. The experimental geometry and crystalline quality of the sample play a crucial role since the sample is used as a beam splitter to obtain the dichroic signal. Therefore, understanding the effect of even small deviations on the signal is essential for the practical use of the EMCD technique.

To obtain a dichroic signal, two angle-resolved electron energy-loss spectra are acquired and subtracted from each other, which is referred to as the EMCD signal. The two scattering angles, described by scattering vectors $\mathbf{k}$, are chosen to be symmetric around a mirror axis \cite{nature}. The sample is often oriented with respect to the incoming electron beam so that mainly two or three spots are visible in the diffraction plane, conventionally named two-beam case (2BC) and three-beam case (3BC) geometries, respectively. Theoretical considerations show that the 3BC is fully symmetric and, according to the sum rules, preferable over the 2BC geometry for quantitative measurements \cite{jantheory}. However, very small deviations from the perfect 3BC geometry are found to influence the extracted dichroic signal \cite{jantheory}. 

In our recent publication, Ref.~\cite{lidbaum}, we presented quantitative measurement of the orbital to spin magnetic moment ratio for a bcc Fe sample. Energy-resolved reciprocal space maps were acquired in 3BC geometry, where we extracted the ratio of the orbital to spin magnetic moment ($m_L/m_S$) and compared with theoretical calculations. Here we build on the findings in Ref.~\cite{lidbaum} and systematically evaluate the influence of the properties of the sample (Sec.~\ref{sec:sample}), the experimental set-up (Sec.~\ref{sec:micSettings}), data treatment (Sec.~\ref{sec:experimentRealReciprocalmaps}) and considerations on the symmetry for the 3BC geometry (Sec.~\ref{sec:results-reciprocalspacemaps}). We also describe the special setting of the microscope which allow us to acquire real space maps of the EMCD signal (Sec.~\ref{sec:micSettings}). Sec.~\ref{sec:theory} describes the theoretical approach applied in our \emph{ab initio} simulations of the EMCD effect. Sec.~\ref{sec:results-reciprocalspacemaps} summarizes our findings, comparing in detail the results from experiment and theory of the reciprocal space maps. We analyze in particular the systematic errors of the observed $m_L/m_S$ ratios caused by deviation from the symmetric geometries, as required by the sum rules. The results of the real space maps are described in Sec.~\ref{sec:results-realspacemaps} and finally followed by the main conclusions in Sec.~\ref{sec:Conclusions}.

%*******************************************************************
%*******************************************************************
%************          SAMPLE PREPARATION           ****************
%*******************************************************************
\section{Sample preparation\label{sec:sample}}
\begin{figure}
  \includegraphics[width=8.5cm]{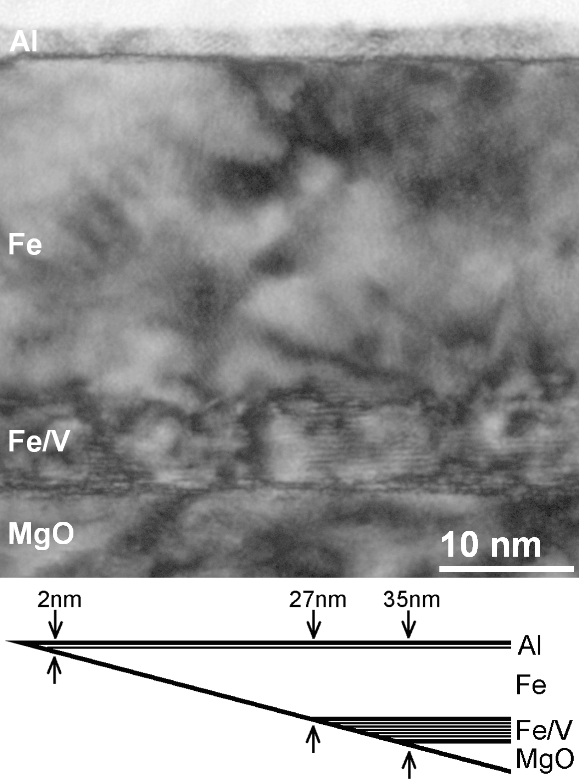}
  \hspace{16pt}
  \caption{A bright field TEM micrograph of the studied Fe sample shown in cross section. MgO substrate, Fe/V super-lattice (to reduce misfit), the Fe layer studied by EMCD (approximately 25 nm) and Al capping as indicated (epoxy glue from TEM sample preparation at top). To enhance the contrast in the image, a gamma intensity transformation function was used. A schematic picture of the sample in cross section is shown in the bottom part of the figure, indicating its total thickness. The EMCD experiments were here performed on a TEM sample prepared in top view geometry.}
\label{fig:sample}
\end{figure}

An 25 nm thick bcc Fe (001) layer was grown by UHV magnetron sputtering on a MgO (001) substrate, using a Fe/V superlattice as a buffer layer. After the substrate was outgassed at 780$^\circ$C in UHV, the temperature was lowered to the deposition temperature of approximately 330$^\circ$C. The seed layer employed was a two step superlattice (4 x [0.91 nm V / 0.29 nm Fe] followed by 4 x [0.45 nm V + 0.29 nm Fe]). As process gas Ar with a purity of 99.9999\% at a pressure of 0.8 Pa was used. Deposition rates for Fe and V were 0.052 nm/s and 0.032 nm/s, respectively. The final Fe layer was protected with a 2 nm thick Al capping layer to prevent oxidation. To study the quality of the grown Fe layer, x-ray diffraction and reflectivity measurements were performed using a Siemens D5000 diffractometer in Bragg-Brentano focusing geometry with Cu K$\alpha$ radiation ($\lambda = 0.154$ nm) and a secondary monochromator. Rocking curve measurements on the Fe (002) peak revealed a texture angle, full width at half maximum (FWHM) of $\sim 0.67^\circ$, reflecting the high quality of the sputtered thin layer. In the thin film community such high quality Fe sample is often referred to as monocrystalline. Though, for the EMCD measurements this texture angle must be taken into account. The difference in electron diffraction of crystalline regions is seen as contrast changes in the Fe layer in Fig.~\ref{fig:sample}, where a bright field TEM image of the sample in cross section is shown.

Conventional sample preparation including dimple grinding and Ar-ion bombardment \cite{alani} from the substrate side in a Gatan Precision Ion Polishing System (PIPS) was used (4 kV, 2.5 kV and finally at 1.3 kV) for the preparation of the top view TEM sample used in the EMCD measurements. Propylene glycol (water content $<0.1\%$) was used during the grinding processes as water based solvents often lead to cracks in MgO substrates (probably because of formation of magnesium hydroxide). The ion milling process was stopped when the size of the hole in the sample was slightly below 5~$\mu$m in diameter. At these conditions, we obtain a region of interest several micrometers wide, where the sample thickness and orientation change little as function of position. A schematic of the sample is shown Fig.~\ref{fig:sample}, together with its total thickness at different positions. A total handling time of about 10 minutes in air was required to transfer the sample from the vacuum chamber of the PIPS into the microscope.

%*******************************************************************
%*******************************************************************
%************          MICROSCOPE SETTINGS          ****************
%*******************************************************************
\section{Choice of microscope settings for signal optimization\label{sec:micSettings}}
The microscope used in this article, and many of today's modern microscopes, use a Schottky emitter as electron source. These emitters typically show a feature called side-lobe emission \cite{scott} that must be considered when measuring the weak signals related to EMCD. Its influence on the EMCD measurements is discussed in this section.

\subsection{Removal of side lobes using telefocus mode\label{sec:ExpSettingsTelefocus}}

\begin{figure}
  \includegraphics[width=10cm]{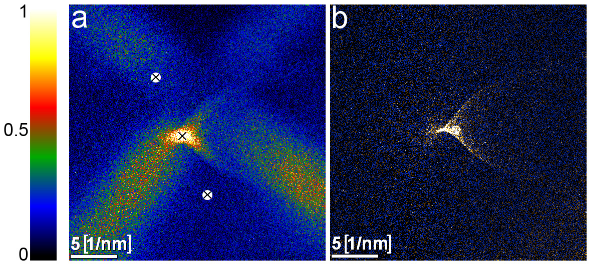}
  \caption {(color online) a) Energy filtered diffraction pattern recorded at 700 eV energy-loss and 20 eV slit using standard gun settings, showing typical side lobe emission. Positions for the Fe $\mathbf{G}$ = ($\pm$200) diffraction spots, used in the EMCD experiment of Fig.~\ref{fig:dich-maps}, are indicated with crossed circles. In b) the gun settings were changed to enter the so-called telefocus mode where the side-lobe emission can be strongly reduced by using the condenser apertures. Both diffraction patterns were normalized by averaging the intensity at the position of the transmitted beam.}
\label{fig:EFDIF-sidelobes}
\end{figure}

The TEM experiments were performed using a FEI Tecnai F-30ST FEG (Schottky emitter) operated at 300 kV, equipped with a post-column spectrometer, Gatan image filter GIF2002. Side-lobe emission occurs when electrons are emitted from not only the tip of the Tungsten filament but also from the relatively large faces of the emitter crystal. The electrons are emitted from the faces with different angles and can therefore also hit other parts of the gun system, for example the extractor anode plate. Thereafter they can be backscattered or cause emission of secondary electrons \cite{scott,TecnaiHelp}. These electrons will be focused differently than electrons of the primary beam and result in a non uniform intensity distribution, observable in the diffraction plane. Energy filtered diffraction (EFDIF) patterns without a sample in the electron beam reveals the distribution of the electrons in reciprocal space as function of the energy-loss. The appearance of the pattern will depend on the settings of the electron gun, the shape of the filament and its position inside the microscope column, and also on the energy of the electrons. An EFDIF pattern from the microscope used here is shown in Fig.~\ref{fig:EFDIF-sidelobes}a, acquired at 700 eV energy-loss with a 20 eV wide energy selecting slit and no sample in the path of the electron beam. Typical standard gun lens settings were used, 4500 V extraction voltage and electrostatic gun lens value of 2 which corresponds to 1000 V in gun lens voltage \cite{TecnaiHelp}. A non-symmetrical pattern is observed in Fig.~\ref{fig:EFDIF-sidelobes}a, which has an intensity of 40-50$\%$ of the transmitted beam intensity in regions where we typically would measure the EMCD signal. This intensity should therefore be considered in EMCD measurements when using Schottky emitters, since it would superimpose onto the magnetic signal, and thereby distorting it. The effect of the side lobes has previously been considered when acquiring diffraction patterns from samples with a very large spacing (such as in biological specimens \cite{TecnaiHelp}), resulting in a hampered visibility of the diffraction spots that are close to the transmitted beam. It should be noted that the use of a small selected area diffraction aperture has been found to remove a significant part of the side lobe emission, although not completely \cite{scott}.

There are two ways of addressing this problem \cite{TecnaiHelp}: {\itshape i)} one can record reference EFDIF patterns (without sample, such as the one in Fig.~\ref{fig:EFDIF-sidelobes}a) with exactly the same energy range as used in the EMCD experiment; the reference patterns are then subtracted from the EMCD data to remove the side-lobe emission. Preferably {\itshape ii)} one can operate the microscope in the so-called telefocus mode where the side-lobe emission is strongly reduced. In telefocus mode the gun settings are changed so that both condenser apertures can be used for removing the side lobes. For the Tecnai microscope used here the extraction voltage was increased to 4550 V in combination with setting the electrostatic gun lens to 5 (corresponding to 500 V in gun lens voltage \cite{TecnaiHelp}), which changes the position of the first cross over after the gun. At these settings the beam will go through a cross over point which is close to the plane of the first condenser aperture. The side lobes can successfully be removed by inserting a small first condenser aperture, while a small second condenser aperture is used for probe forming. The corresponding reference EFDIF pattern is shown in Fig.~\ref{fig:EFDIF-sidelobes}b, revealing the efficient removal of the side lobes. Another consequence of using the telefocus mode is that the intensity of the electron beam is strongly increased, which is useful for the EMCD measurements as it can improve the signal to noise ratio (S/N). Unfortunately, this also makes the alignment procedure more difficult; the intensity of the primary beam can not be changed as normally 'on the fly' by changing spot size (strength of first condenser lens), because the gun tilt settings have to be tuned for each spot size. The high intensity of the electron beam makes it impossible to obtain a spectrum (not in EELS or EFDIF mode) of the zero-loss peak unless the sample is used to attenuate the electron beam (too high beam intensity can saturate and possibly damage this type of CCD camera). This complicates the alignment procedure of the spectrometer. Here the GIF was aligned when in normal gun operation using its built-in software routine, having an image at the viewing screen and no sample in the path of the beam. Thereafter the gun settings were changed to enter telefocus mode and the automatic GIF alignment was repeated, but now with a thick part of the sample introduced in the path of the beam to reduce the intensity on the CCD camera. Typically during the second iteration of the alignment routine only small changes were applied to the settings and the non-isochromatic surface of the GIF was recorded separately (although still with the sample in the electron beam) to make sure that the alignment procedure was successful. Another side effect of using the telefocus mode is that the energy spread of the primary electrons will be slightly increased due to the Coulomb interaction between the electrons in the beam (Boersch effect). This also causes an increase of the minimum probe size due to the position of the cross-over \cite{TecnaiHelp}.

\subsection{Real space maps of the EMCD signal in telefocus mode\label{sec:ExpSettingsRSM}}
One of the novelties promised by the EMCD technique is to obtain information about the magnetic properties of a sample with excellent spatial resolution. An impressive start for high resolution mapping of the magnetic properties was already obtained by using Scanning TEM (STEM) techniques for EMCD \cite{2nm}. The electron beam is focused to a nanometer-sized spot and is scanned over the sample; the electrons scattered at high angles are recorded by a dark field detector to form an image, whereas the electrons scattered at low angles can be collected by the GIF to record the EMCD signal. In theory it is therefore possible to create a real space image of the dichroic signal pixel by pixel. In practice, a much longer acquisition time (10-30 seconds) is needed to collect a single EELS spectrum with reasonable S/N ratio, compared to the few milliseconds usually needed for each pixel in a STEM image. This makes it very difficult to obtain real space maps with a scanning technique because of specimen drift. It is however feasible to first acquire a normal STEM image and use it to select where to record two sets of EELS line scans; from those the EMCD signal can be obtained with high spatial resolution. This method has been used for obtaining an EMCD line scan with a spatial resolution of 2 nm \cite{2nm}. An alternative approach is to use the energy filter with its energy selecting slit to record energy filtered TEM images (EFTEM) from a large part of the sample at once. It has previously been suggested and tested for EMCD but without success \cite{thesisStefano}; the intensity of the electron beam had been too low even for long acquisition times (several minutes). The variations of the dark count of the CCD camera were greater than the eventual dichroic signal, so no meaningful data could be obtained. However, real space maps now become possible due to the increased intensity of the telefocus mode, as will be shown in Sec.~\ref{sec:results-realspacemaps}.

%**************************************************************
%***************RECIPROCAL AND REAL SPACE MAPS*****************
%**************************************************************
\section{Acquisition conditions and data treatment of reciprocal and real space maps\label{sec:experimentRealReciprocalmaps}}

In this section the experimental set-up and data treatment used for reciprocal and real space maps is described in detail. The same procedure was used in Ref.~\cite{lidbaum} for analysing reciprocal space maps.

\subsection{Reciprocal space maps\label{sec:experimentReciprocalMaps}}
The top view sample preparation approach allowed us to choose a region on the sample with appropriate thickness, see Fig.~\ref{fig:sample}. By collecting an electron energy-loss spectrum where a V signal starts to appear (Fe layer thickness 25~nm), it is possible to experimentally calibrate the total mean free path of the inelastically scattered electrons~\cite{egerton}. The calibration is used to obtain the absolute thickness of the probed area using thickness maps. The sample was analyzed at places where the electrons transit only through the pure Fe layer (with Al-capping), at a thickness of $t_{TEM}=19 \pm 2$ nm. Images of the illumination spot on the sample were acquired before and after the series to verify that the sample did not drift during acquisition and to determine the size of the illuminated region. A relatively large beam size was chosen (approximately 600 nm), averaging over many slightly misoriented grains. Series of EFDIF patterns were acquired over the Fe L$_{2,3}$ edges from 750 eV to 670 eV using a slit width of 2 eV and a step size of 1 eV. An exposure time of 5 seconds per diffraction pattern (hardware binning of 2) with a 3.0 mm spectrometer entrance aperture was used. A total acquisition time of approximately 20~minutes was required. The sample was tilted approximately $10^\circ$ from the high symmetry [001] orientation in order to reach the 3BC geometry.

Three-dimensional data sets, so called data cubes \cite{datacube}, are constructed from the acquired EFDIF patterns and represent the fast electron beam intensity as function of energy-loss and scattering vector: $I(k_x,k_y,E)$. The experimental data cubes consist of the ($k_x,k_y$) diffraction planes stacked according to the energy-loss of the fast electrons, which represents the third dimension. EELS spectra can easily be extracted from a data cube by selecting the same pixel or group of pixels in each EFDIF plane and plotting the beam intensity for those pixels as function of the energy-loss. The resulting spectrum is analogous to what would have been recorded in spectroscopy mode by placing a properly shaped aperture at that particular scattering vector ($k_x,k_y$) and using an energy dispersion of 1 eV/channel. The advantage is that the (virtual) aperture size, shape and positions can be optimized \emph{a posteriori}.

To analyze the EMCD signal, the data cube was treated in the following way: a point blemish function removes bright spots arising from spurious radiation; small shifts between the patterns are corrected with sub-pixel precision by cross-correlation \cite{schaffer}; the diffraction patterns are re-binned to 256 x 256 pixels; then, the background under the Fe edge is subtracted by extrapolating from a fit by a power-law model \cite{egerton} to the intensity in the pre-edge region. The point blemish removal was performed by calculating the mean value and standard deviation of the surroundings of every pixel. If the value of a pixel differed from the mean more than a pre-defined multiple of the standard deviation (typical values were from 50 to 100; decided by inspection of the raw data), the pixel was replaced by the mean value. In normal EFTEM series, cross-correlation is needed to compensate for the sample drift during acquisition. In the case of EFDIF series, sample drift does not cause a drift of the diffraction pattern. However, as the primary electron energy is slowly changed to perform the EFDIF series, the software automatically changes the microscope settings to keep the illumination stable by adjusting to the change in accelerating voltage. Small fluctuation in the beam, lenses or software compensation may lead to a drift of the diffraction pattern. The cross-correlation routine is therefore applied by selecting only the transmitted beam to align all EFDIF patterns (typically shifts of less than three pixels within a cube). 

Due to the tilt of the sample from a high symmetry orientation and by uneven scattering from, \textit{e.g.}, the cap layer, the background intensity may vary throughout the ($k_x,k_y$) plane. It is therefore also important to normalize the data cube post-edge to obtain a correct EMCD signal and consequently a correct $m_L/m_S$ ratio. Here, the data cube was normalized by extracting three diffraction patterns in the smooth post-edge region. An average normalization image was calculated and thereafter all diffraction patterns in this data cube were divided by it. The spectra in each pixel of the ($k_x,k_y$) plane were then fitted using the Fe L edge model described in Ref.~\cite{wang}. To model the $L_{2,3}$ peaks a sum of Gaussian/ Lorentzian functions, equal for both peaks and with an asymmetry factor \cite{hesse} was used. For the continuum transition background two arctangent step functions were used, having their respective inflection point at the same position in energy as the $L_{2,3}$ peak and with an intensity ratio of 1:2 \cite{wang}. The steepness of the arctangent function was assumed to be the same at both edges and treated as a fit parameter. The intensity of the resulting peak fitted curve is therefore given by
\begin{eqnarray} \label{eq:fitcurve}
 I_{fit}(E)=\sum_{2,3}~( \underbrace{h_{2,3}A\left\{1+\left[\frac{E-E_{2,3}}{\beta+\alpha\left(E-E_{2,3}\right)}\right]^2\right\}^{-1}}_{Lorenzian}+\underbrace{h_{2,3}(1-A)exp\left\{-\ln(2)\left[\frac{E-E_{2,3}}{\beta+\alpha\left(E-E_{2,3}\right)}\right]^2\right\}}_{Gaussian} \nonumber \\
 +\underbrace{B_{2,3}\left\{\frac{1}{2}+\frac{1}{\pi}\arctan\left[\frac{E-E_{2,3}}{\gamma/2}\right]\right\}}_{core-to-continuum~background}~)~~~~~~~~~~~~~~~~~~~~~~~~~~~~~~~~~~~~~~~~~~~~~
\end{eqnarray}

where $h_{2,3}$ and $E_{2,3}$ are the amplitude and position in energy of the L$_{2,3}$ peaks, respectively. The Lorentzian-Gaussian mixing ratio (A), asymmetry ($\alpha$) and the width (2$\beta$ FWHM) are assumed to be the same for the L$_{2,3}$ peaks. For the background, $B_{2,3}$ describes its amplitude while $\gamma$ the steepness of the $\arctan$ function. Thus, a total of nine fitting parameters were used here. Since the spectra are fitted independently, each fitting parameter is a function of ($k_x,k_y$) that can be illustrated by a reciprocal space map. 

The experimental dichroic maps are obtained by using the integrated intensity under the fitted $L_{2,3}$ edges and an appropriate mirror axis. This approach considers the entire signal strength in the $L_{2,3}$ edges, it removes the arbitrariness of the energy interval selection \cite{lionelsr} for separating the $L_2$ and $L_3$ components and also reduces the influence of peak shifts (due to the non-isochromaticity of the energy filter) on the EMCD signal. The resulting diffraction patterns from the peak fitting of the $L_{2,3}$ areas were rotated in order to apply a mirror axis for the extraction of the EMCD signal. The angle of rotation was obtained by first adding all diffraction patterns in the cross correlated data cube (typically 80), finding the center of the diffracted spots in this resulting pattern and from there determine the rotation angle. It should be noted that the EMCD signal itself might slightly move the Bragg spots perpendicularly to the systematic row. Our calculations have shown that in this experimental set-up, the positions of the diffracted spots are found to move less than 0.01$G_{200}$ for sample thicknesses above 20 nm. Only for specimen thicknesses below 10 nm the spots move at most 0.02$G_{200}$, therefore neglected here.

It should be noted that the $m_L/m_S$ ratio for a sample is a material property, independent of diffraction geometry. 
Nevertheless, in experimental setups the assumptions of sum rules about symmetrical detector positions \cite{oursr,lionelsr} are often not exactly fulfilled, either due to inherent asymmetry of 2BC \cite{janconf2BC} or due to misalignments in orienting the sample to 3BC geometry. Therefore the value of the $m_L/m_S$ ratio extracted by direct application of the sum rules can deviate from the value of the material property, and can also vary throughout the diffraction plane. To keep this distinction - material property vs. value obtained by sum rules - in the further text we refer to the latter as \emph{apparent} $m_L/m_S$ ratio. First principles simulations provide information, how much the apparent $m_L/m_S$ ratio differs from the material property (see Sec.~\ref{sec:theory-SR-evalmLmS}). By using reciprocal space maps we are able to evaluate how the apparent $m_L/m_S$ ratio varies due to asymmetry and noise with ($k_x,k_y$). The $m_L/m_S$ ratio can, according to the sum rules \cite{oursr,lionelsr}, be calculated from:
\begin{equation} \label{eq:ls}
  \frac{m_L}{m_S} = \frac{2}{3} \frac{\int_{L_3} \Delta I(E) dE + \int_{L_2} \Delta I(E) dE}{\int_{L_3} \Delta I(E) dE - 2\int_{L_2} \Delta I(E) dE}
\end{equation}
where $\Delta I(E)$ is the difference between the EELS-spectra extracted at the two specific detector positions with opposite magnetic contribution as defined by the selected mirror axis. A map of the apparent $m_L/m_S$ ratio is thus obtained by applying Eq.~\ref{eq:ls} to the peak fitted data cube. The reciprocal space maps allow for optimization of the position, shape and size of the extraction region in reciprocal space for obtaining reliable $m_L/m_S$ ratio values, since every pixel of the map acts as an individual measurement. Therefore, the S/N ratio can be optimized by selecting regions in the apparent $m_L/m_S$ ratio maps where the EMCD signal is found to be strongest. The histogram of these $m_L/m_S$ ratios for a specific size of the window in reciprocal space was fitted using a Gaussian function. The size of the bins used in the histogram was optimized to best reflect the distribution of $m_L/m_S$ data according to Ref.~\cite{simazaki}; revealing a typical bin size of approximately 0.1. Although, if the bin size is slightly changed, the resulting $m_L/m_S$ ratio does not show a significant change (less than the given uncertainty $\pm 0.01$). The method of peak fitting the histogram was chosen in order to reduce the influence of outliers and thereby get a more reliable quantification of the maps, compared to, \textit{e.g.}, the mean value. The center of the Gaussian has a fit error of only $\pm 0.001$ in $m_L/m_S$.

It should be noted that this error estimation considers the extraction of values from the obtained $m_L/m_S$ map. However, additional errors are present in the peak fitting process, which is more difficult to assess. More fit parameters do not necessary lead to a more accurate fit, wherefore assumptions and simplifications were applied as described above. When having poor energy resolution (here 1-2 eV) there is a tangible risk of introducing fitting artefacts as a result of too few data points. Particularly, the $m_L/m_S$ ratio appears to be most sensitive to the description of the double-step background. Various types of functions for the continuum transition background are used in XMCD such as arctangent, error function, Voigt or even linear. The origin of the double step background is however questioned, since it does not show with the same strength in calculations, but it is often described to originate from $2p$ to $4s$ transitions. Thus, currently no model exists, neither for EMCD nor XMCD, to accurately predict it. Further studies are therefore needed in order to understand its influence on both techniques. However, as dichroic measurements rely on the difference between spectra, the influence of the background is therefore reduced.

\subsection{Real space maps\label{sec:experimentRealMaps}}

As indicated in Sec.~\ref{sec:micSettings}, the strongly increased intensity of the electron beam in telefocus mode allows for real space maps of the EMCD signal. A region of the sample was oriented in a 3BC geometry, exciting $\mathbf{G}$ = ($\pm$200) reflections in Fe. To have good control over the orientation of the sample, single EFDIF patterns and dark field (DF) images were acquired both before and after the experiment using a 5 eV slit at 710 eV. A 5.5 mrad (semi-angle) objective aperture was inserted and placed in one of the two Thales circle positions, to obtain a chiral DF illumination \cite{thesisStefano}. EFTEM images over the Fe $L_{2,3}$ edges were acquired using a slit of 2 eV and step size of 1 eV (hardware binning of 4, 25~s exposure time per image), producing a first data cube consisting of the image plane and the energy-loss of the fast electrons: $I(x,y,E)$. The objective aperture was then shifted to the second Thales circle position to obtain the spectra with opposite magnetic contribution, by acquiring a second data cube of EFTEM images. A total acquisition time of approximately 40~minutes were required per data cube.

The two data cubes were treated in a similar way as for the EFDIF patterns, for details see Sec.~\ref{sec:experimentReciprocalMaps}. First a point blemish function was used to remove noisy spots from spurious radiation. The images were cross correlated \cite{schaffer} by selecting a characteristic feature on the sample that was visible throughout both cubes. Since the EMCD signal will be obtained by taking the difference between the two data cubes, any drift between them must also be removed. To do this, all images in each individually cross correlated data cube were added together (typically 80 images each) and the two resulting images were cross correlated; the resulting drift correction was then applied to the data cubes. Then, the pre-edge background was fitted and subtracted under the Fe $L_{2,3}$ edges, followed by normalized in the post-edge region and finally peak fitted in the same way as for the EFDIF patterns. The EMCD signal was as for the reciprocal space data obtained by subtracting the areas of the peak fitted Fe $L_{2,3}$ peaks.

%***********************************************
%***********************************************
% ***************THEORY*************************
%***********************************************
\section{Theory and simulations\label{sec:theory}}

The dynamical diffraction effects influence the reciprocal space distribution of the magnetic signal in a non-trivial way. Since the S/N ratio in EMCD experiments is generally relatively low, theoretical input about optimizing the experimental geometry is essential. In this section we describe the method of calculations of the reciprocal space maps of the dichroic signal and influence of deviations from symmetrical scattering conditions on the apparent $m_L/m_S$ ratio.

\subsection{Propagation of fast electrons through the crystalline sample\label{sec:theory-prop-fast-electron}}

The propagation of the incoming fast electrons is solved by dynamical diffraction theory applying the Bloch waves approach. Sample thickness, relative orientation of the incoming beam and detector positions are taken into account. Our simulations also include Debye-Waller factors at room temperature \cite{mohanlal}. For the solution of the incoming/outgoing beam eigenvalue problems we use a basis consisting of approximately 700 plane wave components. From the obtained sets of Bloch waves, products of Bloch coefficients are calculated and sorted in descending order. Taking the largest product as unit $U_0$,  a cut-off parameter P is chosen so that products P times smaller than $U_0$ are neglected \cite{jantheory}. In these calculations P was set to $10^{-4}$, which guarantees the accuracy of calculations at the level of about\ 0.1\%.

\begin{figure}
  \includegraphics[width=10cm]{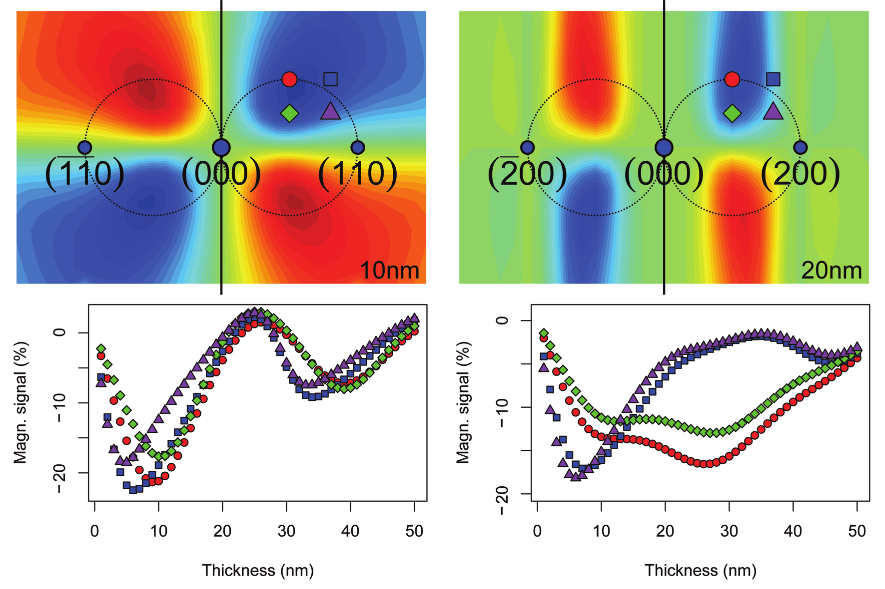}
  \caption{(color online) Upper panels: distribution of the relative dichroic signal (divided by their sum) in the diffraction plane for a perfect 3BC in both $\mathbf{G}=(110)$ and $(200)$ sample orientations (thicknesses as indicated) for the Fe L$_3$ edge, using a vertical mirror axis. Lower panels: thickness dependence at selected points (note the symbol shapes) of the relative dichroic signal, defined as the difference of the integrated $L_3$ edges divided by their sum. 
\label{fig:dich-signal-vs-thickness}}
\end{figure}

Dynamical diffraction effects determine the distribution of the dichroic signal throughout the diffraction plane and its thickness dependence. The energy dependence is then determined by the matrix elements of inelastic transitions (described in Sec.~\ref{sec:theory-inelastic-events}). The strength of the measured dichroic signal depends on all those parameters: sample thickness, composition and orientation as well as the energy of the fast electrons \cite{jantheory}. In Fig.~\ref{fig:dich-signal-vs-thickness} the strength of the dichroic signal, extracted in the vicinity of the Thales circle as a function of sample thickness, is shown. Exciting the bcc Fe $\mathbf{G}=\pm(200)$ reflections instead of the $\mathbf{G}=\pm(110)$ reflections leads to a more uniform dichroic signal (for 300 keV primary beam energy) up to a sample thickness of approximately 30 nm instead of 15 nm, respectively. That was the main motivation for the choice of the $(200)$ systematic row in our experiments.

\begin{figure}
  \includegraphics[width=10cm]{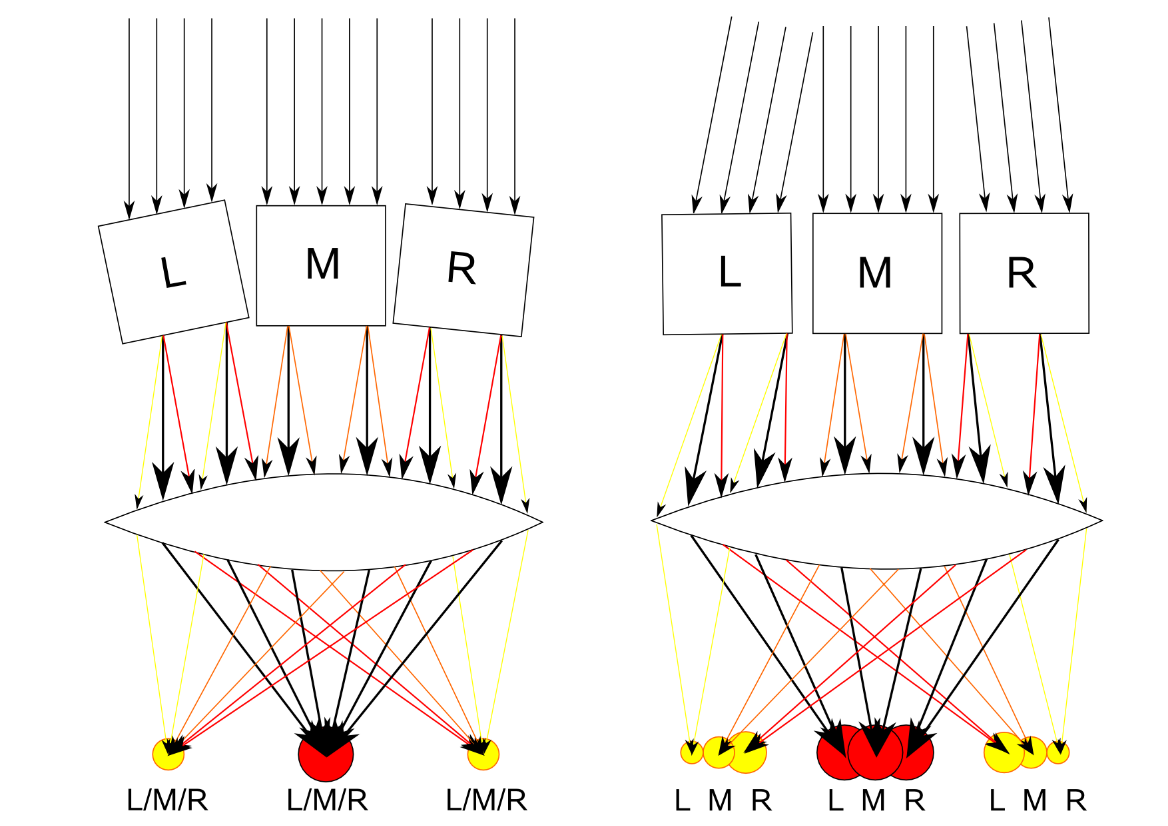}
  \caption{(color online) Scheme of the diffraction on a sample with a texture under parallel illumination conditions (left) and its analogy with a non-parallel beam diffraction on a monocrystal (right). The size of the arrow-heads reflect the strengths of the diffracted beams. See text for details.
\label{fig:drawing2}}
\end{figure}

In order to reproduce the experimental conditions a sample thickness of 19 nm was chosen and to simulate the texture of the sample (the slight misorientations of the crystalline grains), the following procedure was applied (see Fig.~\ref{fig:drawing2}). Small crystallites denoted L (left), M (middle) and R (right) are slightly misoriented, as shown in left part of Fig.~\ref{fig:drawing2}. The parallel illumination then leads to a 3BC for M, but the diffraction conditions for L and R are tilted towards 2BC's of opposite sense, reflected by the size of the arrow-heads of diffracted beams. The lens in the microscope would focus all these contributions into sharp spots, thereby combining all diffraction patterns of individual crystallites (L/M/R). This situation can be equivalently described by a non-parallel electron beam diffracting on an ideal monocrystal (right side of Fig.~\ref{fig:drawing2}). To align all crystallites we need to rotate them, but that needs to be done together with rotation of the partial beams illuminating them. Thus the illumination of the monocrystal is not anymore parallel. We calculate individual diffraction patterns for all incoming beam directions and those lead to the same individual diffraction patterns as before. However, they are now shifted with respect to each other - that is, what normally leads to diffraction disks in convergent electron beam diffraction. To complete the analogy of the crystal with texture, we need to eliminate those shifts, i.e., to make diffraction spots from L, M and R grains overlap again.

To summarize, we describe diffraction of a parallel beam on a sample with a texture angle as diffraction of a non-parallel beam (with the same convergence angle) on a monocrystal, together with a shift of the spots to keep a sharp diffraction pattern. Here we average over 48 incoming beam directions evenly distributed within the illumination cone. Finally, theoretical maps were smeared by convolution with a Gaussian with FWHM of $0.2G_{200}$ (estimated from the width of the transmitted beam in experiment vs. simulation) in order to take into account e.g. aberrations of the spectrometer and point spread of the CCD camera.

\subsection{Description of the inelastic events\label{sec:theory-inelastic-events}}

To describe the inelastic events, knowledge of the initial and final electronic state of an excited atom embedded in a crystalline environment is necessary. This knowledge is obtained here from density functional theory within the local spin density approximation. Relativistic effects are taken into account within the second variational step method setting the magnetization direction to [016] in order to simulate the $10^\circ$ tilt from the high symmetry orientation. The WIEN2k full- potential linearized augmented plane-waves method for the solution of the Kohn-Sham equations was used \cite{wien2k}. More than 100 basis functions per atom and integration over 40000 $\mathbf{k}$-points in the full Brillouin zone ensure a well converged electronic structure. The experimental lattice constant of bulk bcc Fe, $a = 0.2861$ nm, was adopted. To reduce the computational demands we limited the energy integration range to 1 eV, covering the range where the magnetic signal is found to be strongest. Within this energy range the obtained $m_L/m_S$ ratio equals to 0.045, while the ground state value from DFT calculation is 0.021. Details about the formalism and implementation of both dynamical diffraction theory and inelastic event matrix elements can be found in Ref.~\cite{prbtheory}.

\subsection{The sum rules and evaluation of $m_L/m_S$ ratio\label{sec:theory-SR-evalmLmS}}

In XMCD the well-known sum rules \cite{thole,carra} are used to extract spin and orbital moments from observed spectra. These expressions are fundamental for the quantitative analysis of XMCD measurements. For EMCD the sum rules were derived only recently \cite{oursr,lionelsr}, assuming a set of approximations similar to x-ray absorption sum rules, such as taking into account only dipole transitions. In the case of $L_{2,3}$ edges of Fe the sum rules consider transitions from $2p$ to $3d$ states only. In a manner different from XMCD sum rules, the dipole approximation is replaced by the so-called $\lambda = 1$ approximation, which is more accurate for large momentum transfer vectors. This approximation is based on the Rayleigh expansion of the Coulomb interaction term (rather than Taylor expansion) and $\lambda = 1$ denotes the cut-off of this expansion, see Ref.~\cite{prbtheory}. In sum rules, radial wave functions of $2p_{1/2}$ and $2p_{3/2}$ core states are assumed to be the same. Comparing full calculations to calculations with equal core radial wave functions shows that the actual difference between these wave functions reduces the simulated $m_L/m_S$ ratio approximately by 0.01. There is a set of other approximations in the sum rules, detailed in Ref.~\cite{ebert}. The effect of a fully relativistic evaluation of inelastic transition matrix elements is expected to be small for Fe. The exchange splitting of core states in the first order leads to only small shifts of $|jj_z\rangle$ core levels and as such it does not influence energy integrated quantities. The energy dependence of final $3d$ states is taken into account in our simulations but it is neglected in the derivation of the sum rules.

The derivation of the EMCD sum rules is based on the assumption that we take a difference of two spectra measured at symmetrical conditions, which lead to the same set of Bloch waves and Bloch coefficients \cite{oursr}. Under this assumption, the observed spectral differences arise only from the component of the magnetic moment parallel to the beam. As discussed in detail elsewhere \cite{jantheory,janconf2BC}, if this is not accurately fulfilled (for example, if the compared spectra have slightly different dynamical diffraction conditions), the value obtained by application of the sum rules expression contains a systematic error. The apparent $m_L/m_S$ ratio calculated from experimental spectra may substantially differ from the real one if the acquisition geometry is not perfectly symmetric. The studies mentioned above indicate that the 2BC provides good accuracy when measuring inside or close to the Thales circle. The 3BC provides the advantage of two mirror axes, which can be used to form double-difference maps \cite{lidbaum}. This procedure can efficiently minimize the error due to deviation from exact 3BC orientation.

%***********************************************
%***********************************************
%***********RESULTS AND DISCUSSION**************
%***********************************************
\section{Results and discussion\label{sec:results}}
%SHOULD WE ADD AN INTRODUCTION HERE? E.g. In this section the experimental- and theoretical results are described and compared.

\subsection{Reciprocal space maps\label{sec:results-reciprocalspacemaps}}

%3BC vs. 2BC, ADVANTAGES:
The 2BC and 3BC geometries both enable quantitative analysis in EMCD experiments. As previously mentioned, a perfect 3BC is more advantageous than the 2BC geometry as it allows fully symmetric detector positions, which is required by the sum rules for a quantitative analysis \cite{jantheory}. Having reciprocal space maps of the dichroic signal allows for detailed analyses of data, comparing experiment and theory. In this section the quantitative use of the 3BC geometry is described in detail, using the same data set as in Ref.~\cite{lidbaum}. Details on the asymmetry of the 2BC geometry can be found in Refs.~\cite{jantheory} and~\cite{janconf2BC}.

%SHOW EMCD-MAPS:
\begin{figure}
  \includegraphics[width=4.25cm]{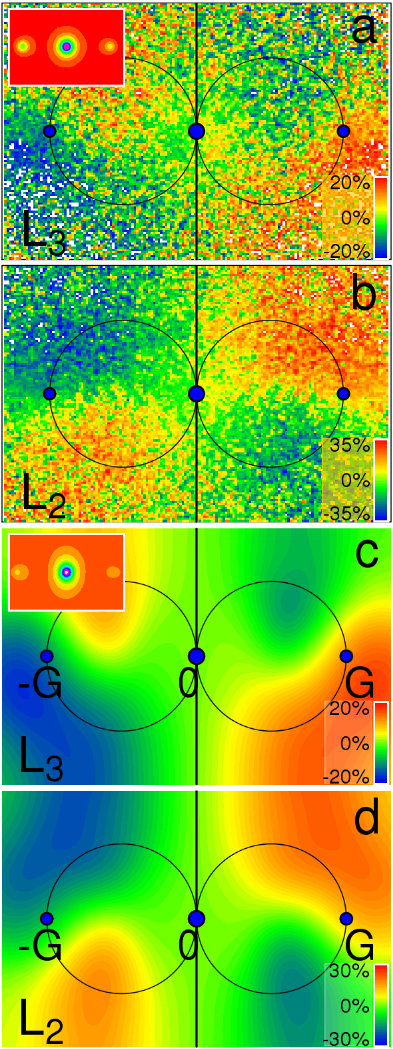}
  \includegraphics[width=4.25cm]{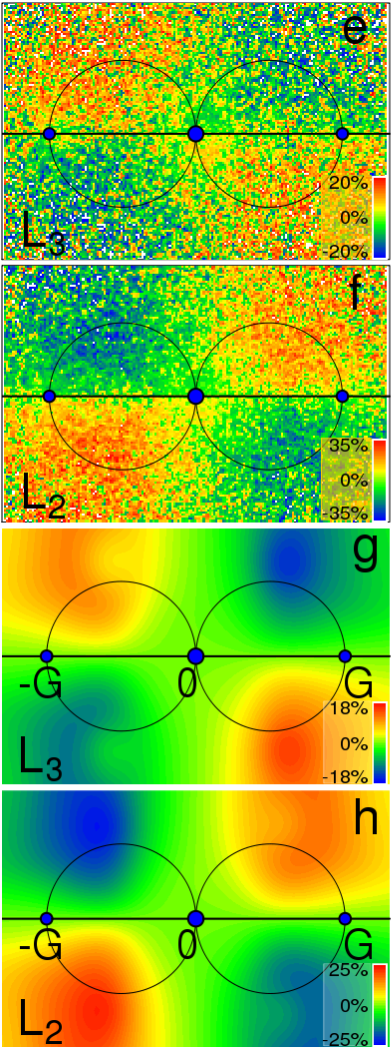}
  \caption{(color online) Reciprocal space maps of the EMCD signal for the Fe $L_{2,3}$ edges with the sample oriented in a 3BC geometry, using a vertical (left column) or horizontal (right column) mirror axis. In a), b) and e), f) experimental maps of relative EMCD signal (difference divided by its sum) at $L_2,3$ edges are shown. The corresponding simulations are shown in c), d) and g), h) considering the texture angle of the sample and the deviation from perfect 3BC geometry as explained in the text. The black line indicates the applied mirror axis and blue spots the positions of the transmitted and Bragg scattered $\mathbf{G} = (200)$ and $-\mathbf{G} = (\bar{2}00)$ beams. The insets show the corresponding experimental and simulated diffraction patterns averaged over an energy interval from 695 eV to 740 eV.} 
  \label{fig:dich-maps}
\end{figure}

To orient our specimen with a texture angle of $\sim 0.67^\circ$ in a perfect 3BC geometry is obviously not possible. Since this is a typical texture angle of heteroepitaxial high-quality sputtered thin films and super-lattice structures, it is important to understand how the dichroic signal is affected. This is discussed in detail in Ref.~\cite{lidbaum}. Here we focus on how a small deviation from the perfect 3BC geometry in the experimental set up will affect the selection of mirror axis and consequently the apparent $m_L/m_S$ ratio. Our simulations have shown that deviations from the perfect 3BC geometry strongly influence the measured dichroic signal. A deviation as small as $0.05\mathbf{G}_{200}$, which is equivalent to a tilt of the sample of $0.02^\circ$, is sufficient to introduce substantial variations of the apparent $m_L/m_S$ ratio (see Sec.~\ref{sec:theory-SR-evalmLmS}) throughout the diffraction plane, particularly in the maps with vertical mirror axis. Therefore, the precision and stability of the sample stage in the TEM is also crucial for orienting the specimen. 

In the inset of Fig.~\ref{fig:dich-maps}a the experimental diffraction pattern averaged over an energy interval from 695 eV to 740 eV is shown. From the measured experimental intensities of $\mathbf{G}$ and $-\mathbf{G}$ diffracted spots it can be seen that the Laue circle center has a slight negative $k_x$-value, \textit{i.e.} the intensity of $-\mathbf{G}$ is higher than $\mathbf{G}$. Here, the deviation from perfect 3BC geometry is found to be approximately $-0.05\mathbf{G}_{200}$. Experimental and calculated maps of the relative dichroic signal (difference divided by its sum) for the $L_{2,3}$ edges are shown in Fig.~\ref{fig:dich-maps}, applying vertical and horizontal mirror axes. The dichroic maps were obtained as described in Sec.~\ref{sec:experimentReciprocalMaps} and~\ref{sec:theory}.

\begin{figure}
  \includegraphics[width=8.5cm]{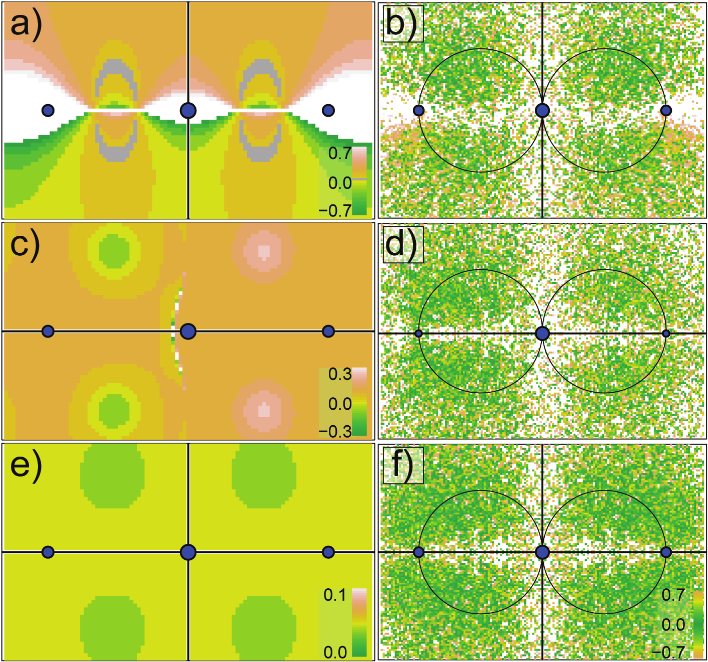}
  \caption{(color online) Calculated (left column) and experimental (right column) reciprocal space maps of the apparent $m_L/m_S$ ratio, using the dichroic maps shown in Fig.~\ref{fig:dich-maps}. The maps are obtained utilizing different mirror axes; a) and b) vertical, c) and d) horizontal and e) and f) double difference (using both). The black lines indicate the mirror axes and blue spots the positions of the transmitted and Bragg scattered $\mathbf{G} = (200)$ and $-\mathbf{G} = (\bar{2}00)$ beams. Note the nearly constant value of the apparent $m_L/m_S$ ratio in the double-difference maps in a large portion of the diffraction plane, despite the texture angle of the sample and the deviation from perfect 3BC geometry. Contrary to that, the use of only one vertical or horizontal mirror axis leads to different values throughout reciprocal space. The color bar of panel f) applies also to b) and d). Values outside the color bar are shown as white.}
  \label{fig:mLdivmS-maps}
\end{figure}

As expected, the distribution of the dichroic signal when using the two different mirror axes shows similar qualitative behavior. Generally, the maps can be split into four quadrants, each with a particular sign of the difference signal. Some differences are visible in the surrounding of the $\pm\mathbf{G}$ diffraction spots when using a vertical mirror axis, see maps in Fig.~\ref{fig:dich-maps}a to d, where the difference signal does not change sign between $L_2$ and $L_3$ edges. This feature is visible in both experiment and calculation and it is caused by the asymmetry introduced when having a slightly misoriented 3BC geometry. Because of that reason, the $m_L/m_S$ ratios extracted at these positions can have a large error. In general, there is a very good correspondence between the theoretical and experimental maps, regarding both position and strength of the maxima and minima of the dichroic signal. In maps with vertical mirror axis the asymmetry is substantial, as is best demonstrated by the simulated maps of the apparent $m_L/m_S$ ratio, Fig.~\ref{fig:mLdivmS-maps}a (note the large range of the scale). Only for a few pixels (grey color) the value of the apparent $m_L/m_S$ ratio is close to the one obtained directly from DFT calculations. Therefore maps using vertical mirror axis are not suitable for measurements of the $m_L/m_S$ ratio, unless a highly precise and accurate 3BC orientation (sub-$0.01^\circ$) is reached.

When the horizontal mirror axis is used, the effect of the misorientation on the dichroic signal is much smaller. Since the higher symmetry of the 3BC geometry is now not used, this case is very similar to the 2BC geometry and it is affected by the same asymmetry problem \cite{janconf2BC}. Nonetheless, these maps show a much smaller variation of the apparent $m_L/m_S$ ratio throughout the diffraction plane, see Fig.~\ref{fig:mLdivmS-maps}c. The major features in this map are visible at $\{110\}$ positions (Thales circle positions). Since the Ewald sphere for incoming beam is closer to spots with positive $k_y$ coordinates, the $(110)$ and $(\bar{1}10)$ spots are excited slightly stronger than their $(1\bar{1}0)$ and $(\bar{1}\bar{1}0)$ counterparts. We note, that here we are working with differences of small numbers. So despite the fact that these excitations are not visible in the diffraction patterns, there are actually slight enhancements of the scattering cross-sections present, and they can become important in difference maps. This effect is well visible in the maps of the apparent $m_L/m_S$ ratio (Fig.~\ref{fig:mLdivmS-maps}c and d) and it is responsible for decreasing the extracted $m_L/m_S$ ratio in the left half-plane, while increasing it in the right half-plane, respectively (see also Table~\ref{tab:table_mLmSsignal}).

When using the horizontal mirror axis the correct $m_L/m_S$ ratio can be extracted at detector positions where; i) the EMCD signal is strong and  ii) the Ewald sphere of the incoming beam is sufficiently far from excitations outside the systematic row around both ``symmetry-related'' detector positions. In the 2BC geometry these conditions are best fulfilled inside or close to the Thales circle, optimally avoiding the $\{110\}$ positions \cite{janconf2BC}. Also the region closest to the systematic row of reflections and to the vertical line passing through the transmitted beam should be excluded since the EMCD signal is there close to zero.

As we have demonstrated, for a sample that is not oriented in a perfect 3BC geometry the apparent $m_L/m_S$ ratio will vary over the reciprocal plane when using only one (horizontal or vertical) mirror axis. To correct for this we introduced in Ref.~\cite{lidbaum} the procedure of double difference maps, where the dichroic signal is extracted as a difference with respect to both horizontal and vertical mirror axes applied successively. This operation effectively cancels out the non-magnetic part of the signal (originating from real parts of mixed dynamic form factors), which remains after using a vertical mirror axis in a measurement when the sample deviates from perfect 3BC orientation. It is assumed that the contribution is close to symmetric with respect to the horizontal mirror axis and is therefore canceled out using the subsequent horizontal mirror axis. This assumption is verified both by simulations and experiments, as demonstrated by the weakly varying apparent $m_L/m_S$ ratio extracted from double difference maps, see Fig.~\ref{fig:mLdivmS-maps}e and f. Therefore, the double difference allows to extract the $m_L/m_S$ ratio from a larger part of the diffraction plane, compared to single mirror axis approach. The apparent $m_L/m_S$ ratios, extracted using a box of $0.5\mathbf{G}_{200}$ x $0.5\mathbf{G}_{200}$ as described in Ref.~\cite{lidbaum}, are summarized in Table~\ref{tab:table_mLmSsignal} for the 3BC geometry while using different mirror axes.

\begin{table}
	\caption{Extracted $m_L/m_S$ ratios in the 3BC geometry using different mirror axes. The values were extracted from experimental and calculated apparent $m_L/m_S$ ratio maps, while optimizing the position of the extracted region of $0.5\mathbf{G}_{200}$ x $0.5\mathbf{G}_{200}$. The error for the experimental values show the standard error from the histogram of the individual $m_L/m_S$ values in the box~\cite{lidbaum}, while for calculated the standard error obtained from mean value is shown.}

	  \centering
		\begin{tabular}{l l l}
		\hline
	                               Axis      & Exp. $m_L/m_S$                          & Calc. $m_L/m_S$         \\ 
		\hline
		\hline		
		                             Vert.     & 0.22$\pm 0.01$(Up)                      & $0.19\pm 0.21$(Up)      \\
		                                       & 0.02$\pm 0.01$(Dn)                      & $-0.11\pm 0.19$(Dn)     \\
		                             Horiz.    & 0.10$\pm 0.01$(L)                       & $0.003\pm 0.04$(L),     \\
		                                       & 0.12$\pm 0.01$(R)                       & $0.071\pm 0.04$(R)      \\
		                             Double    & 0.08$\pm 0.01$                          & $0.036\pm 0.002$        \\
	%	       \textbf{2BC}         & Horiz.    & 0.09$\pm 0.01$                          & $0.085\pm 0.046$        \\
	%	                            & Double    & 0.0?$\pm 0.0?$                          & $0.056\pm 0.021$        \\
		\hline	
		\end{tabular}
		\label{tab:table_mLmSsignal}
\end{table}

\subsection{Real space maps\label{sec:results-realspacemaps}}

\begin{figure}
	\includegraphics[width=8.5cm]{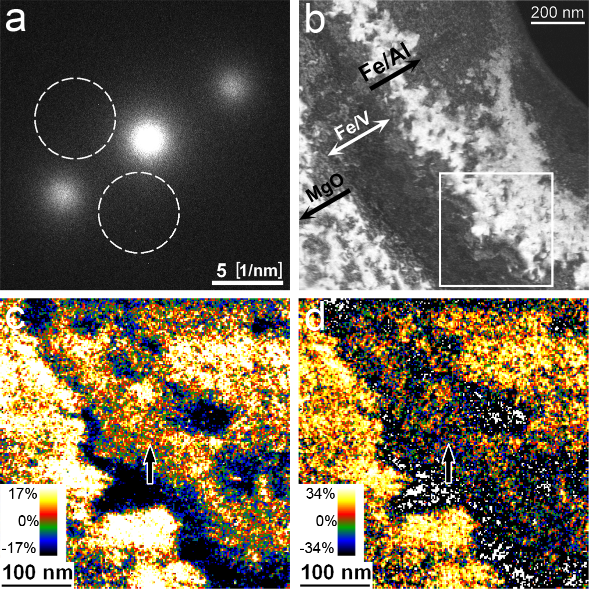}
	\caption{(color online) a) Energy filtered diffraction pattern of the Fe sample at 710 eV using a 5 eV slit, in 3BC geometry with $\mathbf{G}$ = $\pm$(200). The dashed circles indicate the positions of the objective aperture that were used for acquiring the two data cubes. b) Dark field image before the two EFTEM data cubes were acquired, revealing possible regions with correct orientation for the real space maps (bright areas where only Fe). In c) and d) real space maps of the relative EMCD signal (difference divided by its sum) for the $L_{3}$ and $L_{2}$ edges are shown, respectively. The arrows indicate the same positions in c) and d), from where spectra were extracted and is shown in Fig.~\ref{fig:RSMspectra}. The corresponding regions of c) and d) are indicated in b) with a white box.}
\label{fig:RSM}
\end{figure}

Real space maps of the EMCD signal were obtained from the same Fe sample as used for the reciprocal space maps. The sample was oriented in a 3BC geometry to strongly excite the $\mathbf{G} = \pm(200)$ reflections, tilting approximately 20$^\circ$ from the [001] zone axis. We do not expect notable differences of the EMCD signal with respect to the 10$^\circ$ tilt used for reciprocal space maps. An EFDIF pattern at 710 eV using a 5 eV slit was acquired in order to position the objective aperture (diameter of approximately 0.8$\mathbf{G}_{200}$) where we expect a strong EMCD signal, see Fig.~\ref{fig:RSM}a. Although the Fe sample is of high crystalline quality it is (as described in Sec.~\ref{sec:sample}) not a perfect single crystal. Even though the angular deviation measured by XRD is found to be small ($\pm 0.3^\circ$), it is sufficient to show up in a DF TEM image as shown in Fig.~\ref{fig:RSM}b, here selecting the left Fe $\mathbf{G}$ = (200) reflection (Fig.~\ref{fig:RSM}a). Only regions with a bright contrast in this DF image will be relevant for extracting an EMCD signal, here extending over several hundred nanometers. The DF image shows regions with increasing thickness (compare with Fig.~\ref{fig:sample}), starting from a hole in the upper right corner, going through the region of interest with only the Fe layer and Al cap. As the thickness is increased a region with also the Fe/V superlattice occur, where after the MgO substrate is included (lower left corner of Fig.~\ref{fig:RSM}b). 

Two real space data cubes were acquired, one for each of the positions of the objective aperture as indicated in Fig.~\ref{fig:RSM}a. Unfortunately, during acquisition the orientation of the sample relative to the beam changed slightly and thereby further reducing the useful region for the EMCD real space maps. This change could be related to buckling of the sample during electron beam illumination. Suitable regions were found by observing all EFTEM images (in this case 160) throughout both cubes and DF images both before and after the EFTEM series. The data cubes were processed, as described in Sec.~\ref{sec:experimentRealMaps}, to obtain real space maps of the EMCD signal. These maps are shown for the Fe $L_{3}$ and $L_{2}$ edges in Fig.~\ref{fig:RSM}c and d, respectively. A region where the sign of the EMCD signal changes when comparing the $L_2$ to the $L_3$ edge is indicated by the arrow, extending as an approximately 100 nm wide region along the vicinity of the Fe/V superlattice. It should be noted that the EMCD signal outside this region does not change sign between the edges, which becomes evident when studying the EFTEM images of the data cubes. The sample is there found to change its orientation, resulting in strong intensity changes and thus preventing EMCD measurements.

\begin{figure}
	\includegraphics[width=8.5cm]{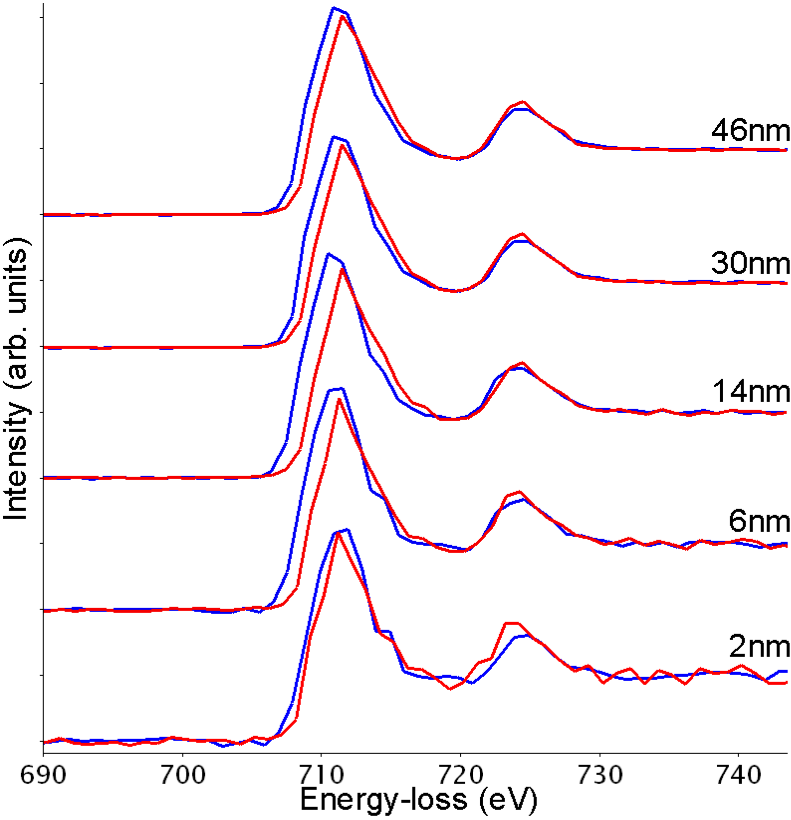}
	\caption{(color online) EELS spectra extracted from the real space data cubes at the position indicated by the arrows in Fig.~\ref{fig:RSM}c) and d). Square boxes with different sizes were used, averaging over 2 nm, 6 nm, 14 nm, 30 nm and 46 nm as indicated. An EMCD signal is observed at both L$_{2,3}$ edges even for the smallest (2 nm) spatial resolution. The spectra were offset in intensity for clarity.}
	\label{fig:RSMspectra}
\end{figure}

Although the maps of the EMCD signal in Fig.~\ref{fig:RSM} are somewhat noisy, they clearly show the possibilities of EFTEM imaging for EMCD. To support the observation, experimental spectra were extracted from the two data cubes at the position of the arrow in Fig.~\ref{fig:RSM}c and d. These are shown in Fig.~\ref{fig:RSMspectra} when selecting different number of pixels by a square box, resulting in a spatial resolution of 2 nm, 6 nm, 14 nm, 30 nm and 46 nm (each pixel corresponds to 2 nm). The spectra were extracted from data cubes that were only processed using point blemish removal and cross correlation. After extraction, the spectra were individually pre-edge background subtracted, aligned in energy-loss and normalized at the post edge region. The results are consistent with the EMCD maps of Fig.~\ref{fig:RSM}, revealing an EMCD signal with a strength of a few percent at the L$_{2,3}$ edges. Thus, this experiment clearly shows the feasibility of EMCD real space maps with nanometer resolution.

Calculating the $m_L/m_S$ ratio is unfortunately not possible with this data set. This could have various sources, such as orientational changes of the sample as discussed above. Other sources of error include the S/N ratio in combination with the poor energy resolution for the peak fitting procedure. It would also be preferable to use four data cubes in order to perform the double difference procedure as described in Sec.~\ref{sec:results-reciprocalspacemaps}, which was not possible here because of sample instability.

%******************************************
%******************************************
%************CONCLUSIONS*******************
%******************************************
\section{Conclusions\label{sec:Conclusions}}
In this article we describe a method for obtaining quantitative measurements of the atom-specific $m_L/m_S$ ratio in the transmission electron microscope using reciprocal space maps. Experiments were performed on a sputtered thin layer of bcc Fe with a small texture angle while deviating from the perfect 3BC geometry, in order to demonstrate the importance of proper mirror axis selection. Even small deviations from a perfect 3BC geometry were found to influence the obtained $m_L/m_S$ ratio when using only horizontal or only vertical mirror axis for extraction of the signal. These deviations can be successfully compensated utilizing both mirror axes, constructing the double difference maps. This is an important step in the development of EMCD for quantitative measurements of technologically interesting samples, such as sputtered thin films and super-lattice structures.

We demonstrate also the importance of removing the so called side-lobe emission, which is observable as a non uniform background in reciprocal space, typically observed for Schottky emitters. It led us to operate the microscope in telefocus mode where, additionally, the intensity of the electron beam is strongly enhanced. The high intensity allowed acquisition of the EMCD signal from EFTEM images, down to the nanometer range. This is therefore an important step forward for the development of the EMCD technique in order to become a quantitative tool for magnetic measurements in the TEM with superior spatial resolution.

%******************************************
%******************************************
%**************ACKNOWLEDGMENTS************
%******************************************
The work was supported through the Swedish Research Council (VR), Knut and Alice Wallenberg foundation (KAW), STINT and computer cluster DORJE of IOP ASCR.

\end{document}